\documentclass[a4paper,twocolumn,11pt,accepted=2021-02-22]{quantumarticle}{}
\pdfoutput=1
\usepackage[utf8]{inputenc}
\usepackage[english]{babel}
\usepackage[T1]{fontenc}
\usepackage{amsmath}
\usepackage{hyperref}
\usepackage{xcolor}

\usepackage{tikz}
\usepackage{lipsum}

\usepackage{mathrsfs}

\newtheorem{definition}{Definition}

\newtheorem{theorem}{Theorem}

\begin{document}

\title{Optimal allocation of quantum resources}

\author{Roberto\ Salazar}
\email{roberto.salazar@ug.edu.pl}
\affiliation{International  Centre  for  Theory  of  Quantum  Technologies,  University  of  Gdansk, Wita Stwosza 63, 80-308 Gdansk,  Poland}
\affiliation{Faculty of Physics, Astronomy and Applied Computer Science, Jagiellonian University, ul. Łojasiewicza 11, 30-348 Kraków, Poland}
\author{Tanmoy\ Biswas}
\affiliation{International  Centre  for  Theory  of  Quantum  Technologies,  University  of  Gdansk, Wita Stwosza 63, 80-308 Gdansk,  Poland}
\affiliation{Institute of Theoretical Physics and Astrophysics, National Quantum
Information Centre, Faculty of Mathematics, Physics and Informatics,
University of Gdansk, Wita Stwosza 57, 80-308 Gdansk, Poland}

\author{Jakub\ Czartowski}
\affiliation{Faculty of Physics, Astronomy and Applied Computer Science, Jagiellonian University, ul. Łojasiewicza 11, 30-348 Kraków, Poland}
\author{Karol\ {\.Z}yczkowski}
\affiliation{Faculty of Physics, Astronomy and Applied Computer Science, Jagiellonian University, ul. Łojasiewicza 11, 30-348 Kraków, Poland}
\affiliation{National Quantum Information
Center of Gdansk, University of Gdansk, ul. Andersa 27 81-824 Sopot, Poland}
\affiliation{Centrum Fizyki Teoretycznej PAN, Al. Lotnik{\'o}w 32/46, 02-668 Warszawa,
Poland}
\author{Pawe\l{}\ Horodecki}
\affiliation{International  Centre  for  Theory  of  Quantum  Technologies,  University  of  Gdansk,  Wita Stwosza 63, 80-308 Gdansk,  Poland}
\affiliation{National Quantum Information
Center of Gdansk, University of Gdansk, ul. Andersa 27 81-824 Sopot, Poland}
\affiliation{Faculty of Applied Physics and Mathematics, National Quantum Information
Centre, Gdansk University of Technology, 80-233 Gdansk, Poland}
\maketitle

\begin{abstract}
The optimal allocation of resources is a crucial task for their efficient use in a wide range of practical applications in science and engineering. This paper investigates the optimal allocation of resources in multipartite quantum systems. 
In particular, we show the relevance of proportional fairness and optimal reliability criteria for the application of quantum resources. Moreover, we present optimal allocation solutions for an arbitrary number of qudits using measurement incompatibility as an exemplary resource theory. Besides, we study the criterion of optimal equitability and demonstrate its relevance to scenarios involving several resource theories such as nonlocality vs local contextuality. 
Finally, we highlight the potential impact of our results for quantum networks and other multi-party quantum information processing, in particular to the future Quantum Internet.
\end{abstract}
\section{Introduction}

In our daily lives, we perform various activities to meet our needs, fulfill our desires or achieve our goals. These activities require the use of physical objects, which may be within our reach or require a provider. Consequently, our physical limitations and the usefulness to perform a specific activity determine an object's value as a resource. Resource theories is a theoretical framework which assigns such a value to an object in a specific physical context  \cite{H4,ResBrandao,ResGen}. Quantum objects are no exception, and at present, their value as resources for different operational tasks has been quantified by resource theories \cite{Barrett,Aolita}.

After establishing an object's value to perform a given task, a significant technical challenge is to distribute available resources to different agents satisfying specific criteria. The solution to the above problem is \emph{the optimal allocation of resources} and defines a well-established research area applied in various fields, including  operations research,  economics, computing, communication networks, and ecology \cite{ResEcono,ResDistri,ResOpt,ResOpt2,ResComp,ResGame,ResNet,ResEcolo}.
However, to date, there has been no systematic research on the optimal allocation of quantum resources.

Online meetings provide a timely example, in which it is possible to understand intuitively the allocation criteria used in this work. We will use this example to present the criteria studied, but for brevity, we will postpone the formal definitions to section 4. In online meetings, it is necessary to distribute the bandwidth fairly among the users (\emph{proportional
fairness}), and we want the communication to be fault-tolerant, i.e. reliable under devices' failures (\emph{reliability}). The network may also impose certain restrictions on the amount of bandwidth allocated to transmitting images or sound that we must consider while carrying out the meeting's tasks in the best possible way (\emph{equitability}). The point raised in this work is that quantum devices' benefits may be subject to the same requirements as classical networks and therefore require optimization under the same criteria.

Precisely, this article investigates the optimal allocation of resources in multipartite quantum systems and illustrates their application in quantum resource theories. Moreover, we show allocations for an arbitrary number of qudits that optimize both the criteria of  \emph{proportional
fairness} and \emph{reliability} for the resource theory of measurement incompatibility. Additionally, we study the \emph{equitability} criteria for resource theories with trade-offs between them and show how this applies to quantum multi-resource scenarios.

\begin{figure}
\begin{centering}
\includegraphics[scale=0.38]{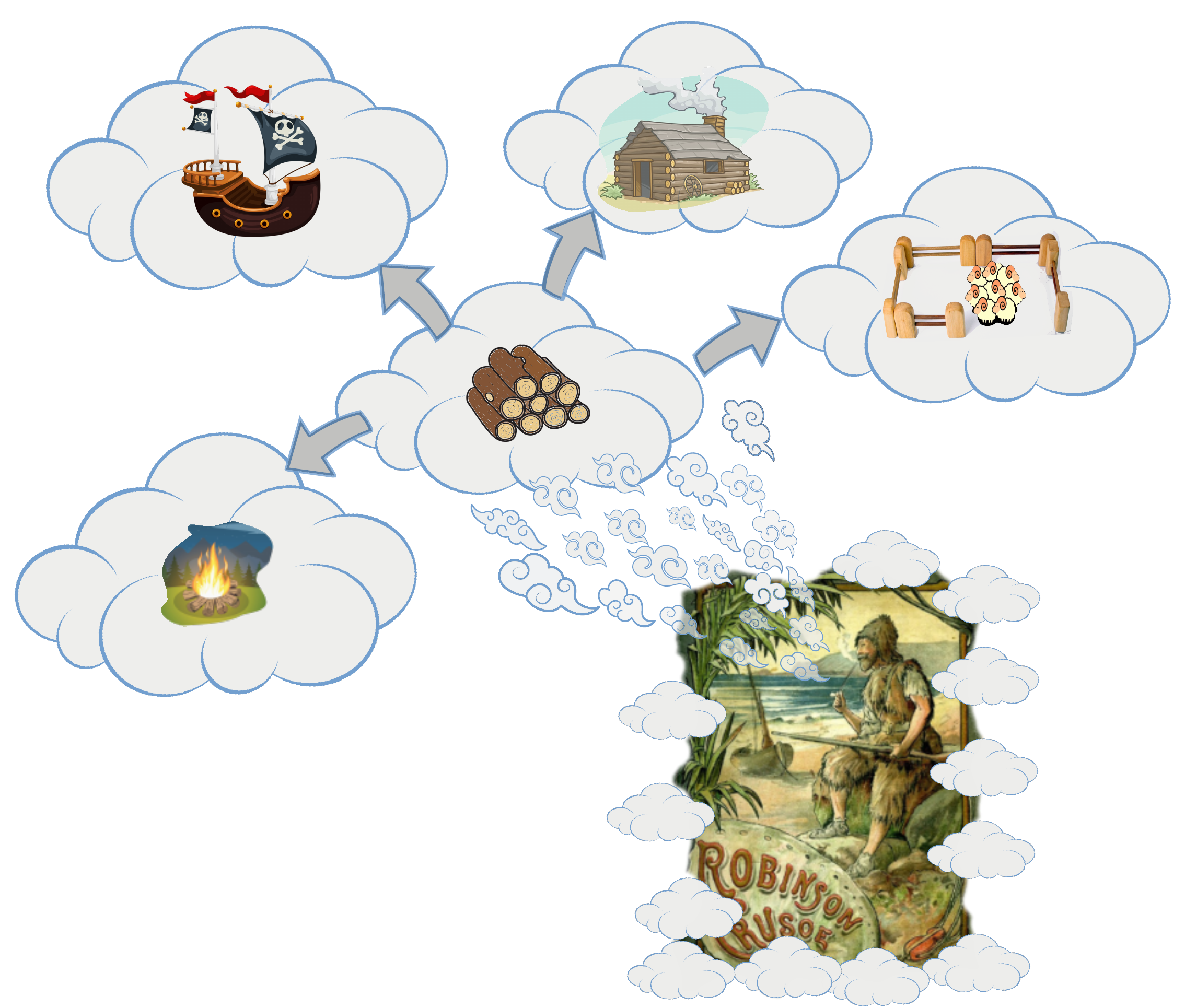}
\par\end{centering}
\caption{\label{fig:Crusoe} A castaway meditates on the optimal allocation of wood logs into several useful tasks such as: prepare a bonfire, build a ship, make a house or provide a fold for his flock.}
\end{figure}
\selectlanguage{english}%


\section{Resource theories in general}

A quantum resource theory's essential idea is to study quantum information processing under a restricted set of physical operations. The permissible operations are called \textit{free} and because they do not cover all physical processes that quantum mechanics allows only certain physically realizable objects of a quantum system can be prepared. Likewise, these accessible objects are called \textit{free}, and any objects that are not free are called a \textit{resource}. Thus a quantum resource theory identifies every physical process as being either free or prohibited and similarly it classifies every quantum object as being either free or a resource. The theory of entanglement forms a representative example of a  quantum resource theory.  For two or more quantum systems, entanglement can be characterized as a resource when free operations are local quantum  operations  and  classical  communication  (LOCC) and free objects are separable states. 

Another important element of a resource theory is a function $\mathcal{M}$
that maps the objects considered by the resource theory into the non-negative
real numbers. The function $\mathcal{M}$ must be non-increasing under
the free operations i.e. a \emph{monotone }and should be proportional
to the advantage of using the object for some operational task. 
\begin{definition}{(Resource theory)}
 A quantum resource theory is defined by a triple $\{\mathcal{F},\mathcal{O} ,\mathcal{M} \}$  where  $\mathcal{F}$ is the set of \textit{free} objects under consideration, for instance quantum states or quantum operations, which  forms a  subset within the set $\mathcal{S}$ of all quantum objects. The set $\mathcal{O}$ of  \textit{free} operations contains functions $o:\mathcal{S}\rightarrow \mathcal{S}$ which preserve the set of free objects. A function $\mathcal{M}:\mathcal{S}\rightarrow [0, \infty +[$, monotone under the set of free operations and such that $\mathcal{M}(f)=0 $ for all $f\in \mathcal{F}$. The function $\mathcal{M}$ is said to measure the resourcefulness of objects in the set $\mathcal{S}$.
\end{definition} 

Because our research focuses on multipartite objects, we assume monotones $\mathcal{M}$ to be well defined when applied to every subsystem. The behaviour on multipartite objects of several standard measures, such as robustness, trace distance and entropic measures is already well known in the literature from the study of convertibility tasks, for more details see \cite{ResGen}.
\medskip{}

In the following we introduce the basic definitions of the resource
theory of incompatibility which recently attracted the attention of the
community \cite{PSkrz,RoB}. We will use this particular theory to
illustrate the application of these definitions and show how they
lead to sound but tractable optimization problems.

\section{Resource theory of incompatibility}

In quantum theory  measurements are described  by Positive Operator Value Measures (POVM). We define a POVM in dimension $d$ and a number of outcomes $n$ denoted by $\textbf{M}:= \{M_1,M_2,\ldots,M_n\}$    where each $M_i \in \mathcal{L}(\mathscr{H}_d)$ is called measurement operator such that $M_i \geq 0$ and $\Sigma_iM_i = \textbf{I}$. Suppose a set of measurements $\{\textbf{M}_x\}_{x}$ where $x= 1;...;m$, each described by measurement operators $M_{a|x}$ labels each of the measurement outcomes $a$ where $a=1,..,o$ such that $\Sigma_{a=1}^{o}M_{a|x}=\textbf{I}$ for every $x$. This set is said to be jointly measurable (or compatible) if there exists a \textit{parent} POVM $\mathcal{G}$ with measurement operators $\mathcal{G}_{\lambda}$ and conditional probability distributions $p(a|x,\lambda)$ such that $ M_{a|x}=\Sigma_{\lambda}p(a|x,\lambda)\mathcal{G}_{\lambda}$. Otherwise the set is said to be incompatible.  One can introduce a resource-theoretic framework in the case of incompatibility with jointly measurable (compatible) POVM as a free measurement set $\mathcal{F}$ and   the set of free operations $\mathcal{O}$ is: any single compatible measurement operator, classical post-processing and random
mixing of single compatible measurements\cite{PSkrz}.

\selectlanguage{english}%
A recent important result shows that for every resourceful measurement
$\mathbf{M}$ there is an instance of minimal-error quantum state discrimination
game for which $\mathbf{M}$ gives greater success probability than
all free measurements \cite{MeasOpT1,MeasOpT2,MeasOpT3}. It has
also been shown that the relative advantage of a resourceful measurement
for state discrimination is proportional to the robustness measure
\cite{Dis7,Dis8}, which quantifies the minimal amount of noise that
has to be added to a POVM to make it free. The formal expression for
the \textit{generalized robustness} of a measurement $\mathbf{M}$ is:
\begin{equation}
\label{eq:robustness}
\mathcal{R}_{g}\left(\mathbf{M}\right)=\min\left\{ s\in\mathbbm{R}_{\geq0}\mid\exists\mathbf{N}\,\,\textrm{s.t.}\,\,\frac{\mathbf{M}+s\mathbf{N}}{1+s}\in\mathcal{F}\right\} 
\end{equation}

This is a monotone measure which means that for any measurement $\mathbf{M}$
and free operation $\phi\in\mathcal{O}$ we have $\mathcal{R}_{g}\left(\phi\left[\mathbf{M}\right]\right)\leq\mathcal{R}_{g}\left(\mathbf{M}\right)$.
We use $\mathcal{R}_{g}$ in our work since it provides a clear
operational interpretation of measurement incompatibility and is easy to define for any subsystem: simply choose any $\mathbf{N}$ that acts on the same systems as $\mathbf{M}$ such that the mixture is some compatible POVM on the corresponding systems.



\section{Allocation of resources}
The allocation of resources consists of \emph{distributing resources into a set of tasks according to a practical criterion}, which usually involves optimizing a figure of merit. When the selected distribution method optimizes the figure of merit in the criteria, we have an \emph{optimal allocation of resources}. We can see in Fig.\ref{fig:Crusoe} a classic example of this problem, where an isolated man on an island must decide the best way to use the available resources.

\begin{figure}
\begin{centering}
\includegraphics[scale=0.7]{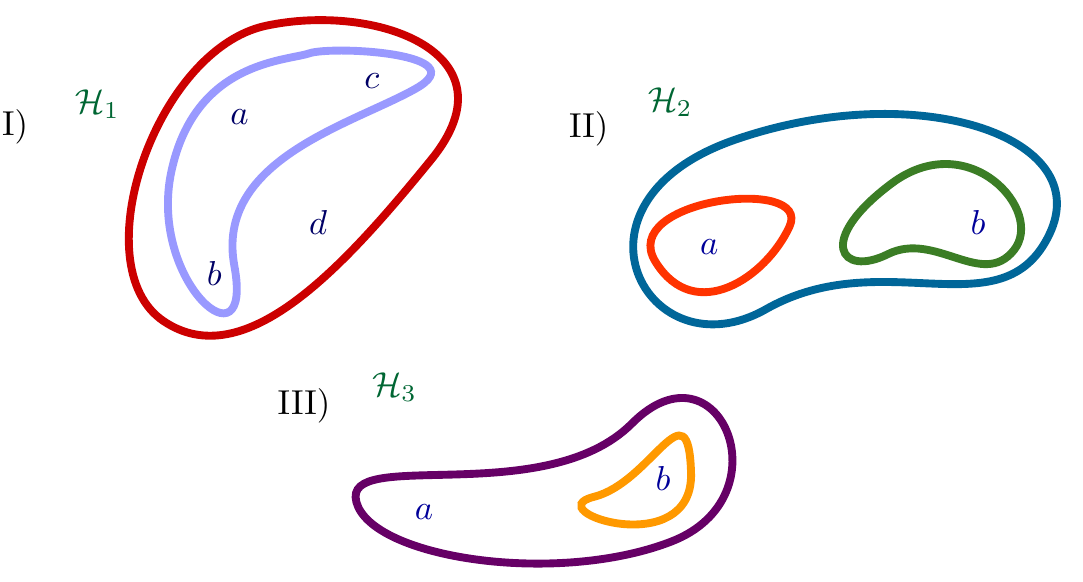}
\par\end{centering}
\caption{\label{fig:hypergraphs}Examples of Hypergraphs used to determine
the relevant parts of the resource allocation. Explicitly, the corresponding
sets of hyperedges are $\mathcal{E}_{\mathcal{H}_{1}}=\left\{ \left\{ a,b,c,d\right\} ,\left\{ a,b,c\right\} \right\} $,
 $\mathcal{E}_{\mathcal{H}_{2}}=\left\{ \left\{ a,b\right\} ,\left\{ a\right\} ,\left\{ b\right\} \right\} $ and $\mathcal{E}_{\mathcal{H}_{3}}=\left\{ \left\{ a,b\right\}  ,\left\{ b\right\} \right\} $.}
\end{figure}

In this article we restrict ourselves to study the above general problem in the case of multipartite quantum objects (a state $\rho$, a measurement
$\mathbf{M}$ or a channel $\Xi$) denoted here by a common symbol $\sigma$. In a multipartite object, some subsets of the systems can be selected for different operational tasks. To describe this selection of relevant sets we shall use the notion of hypergraph $\mathcal{H}=\left\{ \mathcal{V}_{\mathcal{H}},\mathcal{E}_{\mathcal{H}}\right\} $,
in which hyperedges $\alpha\in\mathcal{E}_{\mathcal{H}}$ connect
two or more vertices $v\in\mathcal{V}_{\mathcal{H}}$ \cite{Hyp}.
W\foreignlanguage{english}{e define a \emph{quantum resource allocation}
as the distribution of a quantum multipartite resource $\sigma$ over hypergraph
$\mathcal{H}$ i.e. a list $\mathcal{A}_{\mathcal{H}}\left[\sigma\right]$
of values of the measure $\mathcal{M}$ applied to every reduction
of $\sigma$ to parties in a hyperedge $\alpha$ of $\mathcal{H}$. The way in which the measure is applied for each $\alpha\in\mathcal{E}_{\mathcal{H}}$ must be specified in each case, which we exemplify later in our applications of our framework. \emph{
}To illustrate the concept of resource allocation we present hypergraphs $\mathcal{H}_{1}$, $\mathcal{H}_{2}$ and $\mathcal{H}_{3}$
in Fig.\ref{fig:hypergraphs} which have allocations:
\begin{eqnarray*}
\mathcal{A}_{\mathcal{H}_{1}}\left[\sigma_{abcd}\right] & = & \left\{ \mathcal{M}\left(\sigma_{abcd}\right),\mathcal{M}\left(\textrm{Tr}_{d}\left(\sigma_{abcd}\right)\right)\right\} \\
\mathcal{A}_{\mathcal{H}_{2}}\left[\sigma_{ab}\right] & = & \left\{ \mathcal{M}\left(\sigma_{ab}\right),\mathcal{M}\left(\textrm{Tr}_{a}\left(\sigma_{ab}\right)\right),\mathcal{M}\left(\textrm{Tr}_{b}\left(\sigma_{ab}\right)\right)\right\} \\
\mathcal{A}_{\mathcal{H}_{3}}\left[\sigma_{ab}\right] & = & \left\{ \mathcal{M}\left(\sigma_{ab}\right),\mathcal{M}\left(\textrm{Tr}_{a}\left(\sigma_{ab}\right)\right) \right\}
\end{eqnarray*}
}
\foreignlanguage{english}{If the list $\mathcal{A}_{\mathcal{H}}$
satisfies an \emph{optimality criterion} $\mathscr{C}$ then we will
say that the quantum resource allocation is \emph{optimal.} The relevant
criterion $\mathscr{C}$ depends on the allocation features we wish
to optimize. If a task can be carried out by parts of the entire system
specified by a hypergraph $\mathcal{H}$, then the \emph{performance
}$\Phi_{\mathscr{C}}$ of the allocation is given by a sum over hyperedges
$\alpha\in\mathcal{E}_{\mathcal{H}}$ \cite{ResOpt,ResOpt2}:
\begin{equation}
\Phi_{\mathscr{C}}\left(\mathcal{A}_{\mathcal{H}}\left[\sigma\right]\right)=\sum_{\alpha\in\mathcal{E}_{\mathcal{H}}}\Phi_{\mathscr{C},\alpha}\left(\mathcal{M}\left(\sigma_{\alpha}\right)\right)\label{eq:performance}
\end{equation}
Here $\Phi_{\mathscr{C},\alpha}:\mathbbm{R}_{\geq0}\rightarrow\mathbbm{R}$
is some monotone function of the amount of resource allocated in hyperedge
$\alpha$ and $\sigma_{\alpha}$ is the reduction of $\sigma$ in
which the complement of $\alpha$ has been traced out. The properties
of $\Phi_{\mathscr{C},\alpha}$ depend on how the allocation of resources
at $\alpha$ contribute to the satisfaction of criterion $\mathscr{C}$.
Since the operational task has associated the resource measure $\mathcal{M}$,
the optimization problem consist in finding the multipartite resource
$\sigma$ such that $\mathcal{A}_{\mathcal{H}}\left[\sigma\right]$
maximizes $\Phi_{\mathscr{C}}$ when $\mathcal{H}$ is fixed. The
first optimality criterion $\mathscr{C}_{1}$ we will consider is the
\emph{optimal proportional fairness }criteria.This criterion is required when it is essential to allocate resources fairly among the parties. For example, these include allocation of bandwidth in telecommunication networks, takeoff and landing slots at airports, and water resources.
For this case we maximize the performance of proportional fairness
$\Phi_{\mathscr{C}_{1}}$ given by:
\begin{equation}
\Phi_{\mathscr{C}_{1}}\left(\mathcal{A}_{\mathcal{H}}\left[\sigma\right]\right)=\sum_{\alpha\in\mathcal{E}_{\mathcal{H}}}\log\left(\mathcal{M}\left(\sigma_{\alpha}\right)\right).\label{eq:performance-1}
\end{equation}
Hence, each $\Phi_{\mathscr{C}_{1},\alpha}\left(\cdot\right)$ is
equal to $\log\left(\cdot\right)$. The choice $\Phi_{\mathscr{C}_{1},\alpha}\left(\cdot\right)=\log\left(\cdot\right)$
is useful in the case when no set of parts $\alpha$ has priority
over others for the task performed \cite{ResOpt}. In the case when
certain priority order of the $\alpha\in\mathcal{E}_{\mathcal{H}}$
exists, a different figure of merit is used to ensure equitability
\cite{ResOpt} and we will discuss it later in the article. For example,
the figure of merit $\Phi_{\mathscr{C}_{1}}$ is applied when in a
communication network each demand is routed on a specified single
path. From a resource point of view, the objective is to provide a
resource $\sigma_{\alpha}$ to parties in the set $\alpha$ to communicate
with an additional party $e\notin\mathcal{V}_{\mathcal{H}}$ in a
way as fair as possible among all $\alpha\in\mathcal{E}_{\mathcal{H}}$.
It can be shown that the optimal solution $\mathcal{A}_{\mathcal{H}}\left[\sigma^{\ast}\right]$
satisfies: 
\[
\sum_{\alpha\in\mathcal{E}_{\mathcal{H}}}\left(\frac{\mathcal{M}\left(\sigma_{\alpha}\right)-\mathcal{M}\left(\sigma_{\alpha}^{\ast}\right)}{\mathcal{M}\left(\sigma_{\alpha}\right)}\right)\leq0
\]
for every other $\mathcal{A}_{\mathcal{H}}\left[\sigma\right]$. Precisely, the solution $\mathcal{A}_{\mathcal{H}}\left[\sigma^{\ast}\right]$
is called \emph{proportionally fair} since the aggregate of proportional
changes with respect to any other feasible solution is zero or negative
\cite{ResOpt}.}

\selectlanguage{english}%
Next, we consider the \emph{optimal reliability }criterion $\mathscr{C}_{2}$
which determines an optimal performance
for a task in the presence of failures of the devices used by any
one of its parts in each $\alpha\in\mathcal{E}_{\mathcal{H}}$. The
appropriate performance $\Phi_{\mathscr{C}_{2}}$ which we maximize
in this case reads:
\begin{equation}
\Phi_{\mathscr{C}_{2}}\left(\mathcal{A}_{\mathcal{H}}\left[\sigma\right]\right)=\sum_{\alpha\in\mathcal{E}_{\mathcal{H}}}\pi_{\alpha}\mathcal{M}\left(\sigma_{\alpha}\right),\label{eq:performance-1-1}
\end{equation}
where $\pi_{\alpha}$ denotes the prior probability that only parts
in $\alpha$ will perform the task for which $\sigma$ is the resource.
If every part $\left\{ a_{1},\ldots,a_{n}\right\} $ has a prior chance
$p_{a_{k}}$to work, then: 
\begin{equation}
\pi_{\alpha}=\prod_{a_{i}\in\alpha}p_{a_{i}}\prod_{a_{j}\notin\alpha}\left(1-p_{a_{j}}\right).
\end{equation}
The problem of optimal reliability is a relevant problem for different
engineering areas which is usually addressed by \emph{redundacy models}
\cite{ResOpt2}. In \eqref{eq:performance-1-1} we adapt the classical
performance function to fit the scheme of resource theories (see for
instance the classical definition in section 1.23 in \cite{ResOpt2})
by choosing the $\Phi_{\mathscr{C}_{2},\alpha}\left(\cdot\right)$
to be the probability of successful performace only of parts in $\alpha$
rather than probability of failure, because $\mathcal{M}\left(\sigma_{\alpha}\right)$
is a measure of advantage provided by $\sigma_{\alpha}$ to the corresponding
task.

The previous examples of performance and optimal reliability considered the case of a resource that should be shared in an optimal way among parties performing the same task. 
It should be noted that in principle one can define different tasks for each set $\alpha\in\mathcal{E}_{\mathcal{H}}$ in the two previous allocation problems. However, as we shall see on the following examples, the natural cases that involve multiple tasks are those where the availability of the resource is limited and - what is  even more important - 
 competition between the tasks exist. To describe such a situation in our notation consider for every resource $\sigma$ and all $\alpha\in\mathcal{E}_{\mathcal{H}}$: 
\begin{eqnarray}
\sum_{\alpha\in\mathcal{E}_{\mathcal{H}}}\Lambda_{\alpha}\left(\sigma\right)\mathcal{M}_{\alpha}\left(\sigma_{\alpha}\right) &\leq& \Gamma\left(\sigma\right)  \: \label{eq:tradeoff}\\
L_{\alpha}\leq\mathcal{M}_{\alpha}\left(\sigma_{\alpha}\right) &\leq& U_{\alpha}  \:.\label{eq:limits}
\end{eqnarray}
 with $\Lambda_{\alpha}\left(\sigma\right),\Gamma\left(\sigma\right)$ real non-negative functions of states and $L_{\alpha},U_{\alpha}$ some non-negative constants. Here we use explicitly  the index dependence to express the possibility that the parties in each  $\alpha\in\mathcal{E}_{\mathcal{H}}$ measure the resourcefulness of their $\sigma_{\alpha}$ with a different $\mathcal{M}_{\alpha}$. The choice of each $\mathcal{M}_{\alpha}$ is determined by the task assigned to the resource in each $\alpha$. The equations \eqref{eq:tradeoff}
stand for the trade-offs between the resources assigned for tasks
at different $\alpha$ while the \eqref{eq:limits} determine prior
limitations of the resources for each task. This set of inequalities
is known as\emph{ knapsack constraints} (KC) \cite{ResOpt}. In
the case of two or more tasks involved and non-trivial knapsack constraints
a typical optimality criterion $\mathscr{C}_{3}$ is the \emph{optimal
equitability} which is defined by a recursive max-min algorithm:
\begin{enumerate}
\item If upperbounds $U_{\alpha_{i}}\leq U_{\alpha_{j}}$ then introduce
the condition $\mathcal{M}_{\alpha_{i}}\left(\sigma_{\alpha_{i}}\right)\leq\mathcal{M}_{\alpha_{j}}\left(\sigma_{\alpha_{j}}\right)$.
The corresponding partial order determines the minimization over $\alpha\in\mathcal{E}_{\mathcal{H}}$ in
the following steps.
\item If $\mathcal{\mathcal{E}_{\mathcal{H}}}\neq\left\{ \emptyset\right\} $
then find a solution with prior KC for 
\[
\Phi_{\mathscr{C}_{3}}\left(\mathcal{A}_{\mathcal{H}}\left[\sigma\right]\right)=\sup_{\sigma}\min_{\alpha\in\mathcal{E}_{\mathcal{H}}}\mathcal{M}_{\alpha}\left(\sigma_{\alpha}\right)
\]
\item Let us call $\alpha_{1}$ the $\alpha\in\mathcal{E}_{\mathcal{H}}$
such that $\mathcal{M}_{\alpha_{1}}\left(\sigma_{\alpha_{1}}\right)$
is the minimum on step 1. If $\mathcal{E}_{\mathcal{H}}/\alpha_{1}=\mathcal{E}_{\mathcal{H}}^{(1)}\neq\left\{ \emptyset\right\} $
then find a solution with prior KC for: 
\[
\Phi_{\mathscr{C}_{3}}\left(\mathcal{A}_{\mathcal{H}^{(1)}}\left[\sigma\right]\right)=\sup_{\sigma}\min_{\alpha\in\mathcal{E}_{\mathcal{H}}^{(1)}}\mathcal{M}_{\alpha}\left(\sigma_{\alpha}\right)
\]
under the additional constraint:
\[
\ensuremath{\Phi_{\mathscr{C}_{3}}\left(\mathcal{A}_{\mathcal{H}}\left[\sigma\right]\right)=\mathcal{M}_{\alpha_{1}}\left(\sigma_{\alpha_{1}}\right)\leq\mathcal{M}_{\alpha}\left(\sigma_{\alpha}\right)}
\]
for all $\ensuremath{\alpha\in\mathcal{E}_{\mathcal{H}}^{(1)}}$.
\item Iterate with corresponding $\alpha_{n-1}$ from $\alpha\in\mathcal{E}_{\mathcal{H}}^{(n-2)}$
such that $\mathcal{M}_{\alpha_{n-1}}\left(\sigma_{\alpha_{n-1}}\right)$
is the minimum on step $n-1$. If $\mathcal{E}_{\mathcal{H}}^{(n-2)}/\alpha_{n-1}=\mathcal{\mathcal{E}_{\mathcal{H}}}^{(n-1)}\neq\left\{ \emptyset\right\} $
then find a solution with prior KC for
\[
\Phi_{\mathscr{C}_{3}}\left(\mathcal{A}_{\mathcal{H}^{(n-2)}}\left[\sigma\right]\right)=\sup_{\sigma}\min_{\alpha\in\mathcal{E}_{\mathcal{H}}^{(n-1)}}\mathcal{M}_{\alpha}\left(\sigma_{\alpha}\right)
\]
under the additional constraints:
\[
\mathcal{M}_{\alpha_{1}}\left(\sigma_{\alpha_{1}}\right)\leq...\leq\mathcal{M}_{\alpha_{n-1}}\left(\sigma_{\alpha_{n-1}}\right)\leq\mathcal{M}_{\alpha}\left(\sigma_{\alpha}\right)
\]
for all $\ensuremath{\alpha\in\mathcal{E}_{\mathcal{H}}^{(n-1)}}$
and such that: 
\[\Phi_{\mathscr{C}_{3}}\left(\mathcal{A}_{\mathcal{H}^{\left(m-1\right)}}\left[\sigma\right]\right)=\mathcal{M}_{\alpha_{m}}\left(\sigma_{\alpha_{m}}\right)\]
for each $\mathcal{M}_{\alpha_{m}}\left(\sigma_{\alpha_{m}}\right)$.
\item When $\mathcal{E}_{\mathcal{H}}^{(m-1)}=\left\{ \emptyset\right\}$ 
stop.
\end{enumerate}
In the previous recursive algorithm minimization of $\mathcal{M}_{\alpha}\left(\sigma_{\alpha}\right)$
over $\alpha\in\mathcal{E}_{\mathcal{H}}$ considers $\mathcal{M}_{\alpha_{i}}\left(\sigma_{\alpha_{i}}\right)\leq\mathcal{M}_{\alpha_{j}}\left(\sigma_{\alpha_{j}}\right)$
iff $U_{\alpha_{i}}\leq U_{\alpha_{j}}$ because in this case allocating
resources in $\alpha_{j}$ is potentially more advantageous than in
$\alpha_{i}$. We also note that the above algorithm will halt in
finite number steps because the number of $\alpha\in\mathcal{\mathcal{E}_{\mathcal{H}}}$
is finite. Solutions to optimal equitability are known to
be non-unique, but nevertheless they provide a set of solutions considered
safe for the usual practical applications \cite{ResOpt}. If this
is not satisfactory, a unique solution can be ensured by additional
requirements such as Pareto optimality \cite{ResOpt} of the final
order $\mathcal{M}_{\alpha_{1}}\left(\sigma_{\alpha_{1}}\right)\leq...\leq\mathcal{M}_{\alpha_{N}}\left(\sigma_{\alpha_{N}}\right)$.

It is essential to highlight that the theoretical framework we developed for the allocation of quantum resources is entirely general and that it is possible to apply it in any case that requires at least two agents performing more than one instance of a task or more than one task with their quantum devices. That said, the next section's results illustrate the usefulness of the theoretical framework by demonstrating concrete and relevant applications.

\section{Results}

\subsection{Fair and reliable allocation of incompatible measurements}

Before presenting a specific example of resource allocation let us
recall the notion of unbiased bases. Two orthogonal measurements with
labels $j,j^{\prime}\in\left\{ 0,1\right\} $ are \emph{unbiased } iff
they satisfy the following condition:
\begin{equation}
\left|\left\langle u_{a}^{j}\mid u_{b}^{j^{\prime}}\right\rangle \right|^{2}=\begin{cases}
\delta_{ab} & ;\,j=j^{\prime}\\
\frac{1}{D} & ;\,j\neq j^{\prime}
\end{cases}\label{eq:MUB}
\end{equation}
where $D$ is the total dimension of the system \cite{MUB1,MUB2}.
Two unbiased bases consisting of product states only were introduced
in \cite{MUB3}, for which case we write $\left|u_{\mathbf{a}}^{x}\right\rangle =\left|u_{a_{1}}^{x}\right\rangle \otimes\ldots\otimes\left|u_{a_{N}}^{x}\right\rangle $
with each $\left|u_{a_{k}}^{x}\right\rangle $ a state from two unbiased
bases. Therefore, for $N$ systems of dimension $d$ two product unbiased
measurements $\left|u_{\mathbf{a}}^{x_{1}}\right\rangle \left\langle u_{\mathbf{a}}^{x_{1}}\right|,\left|u_{\mathbf{b}}^{x_{2}}\right\rangle \left\langle u_{\mathbf{b}}^{x_{2}}\right|$
satisfy \eqref{eq:MUB} replacing $\delta_{ab}\rightarrow\delta_{\mathbf{ab}}$
and $D=d^{N}$. It was demonstrated in \cite{MUB3} that one can
find at least two product bases in $\mathscr{H}_{d}^{\otimes N}$
which are mutually unbiased and such that any of their reductions
to $M$ qudits forms a set of two unbiased product bases in $\mathscr{H}_{d}^{\otimes M}$.
The above property demonstrates that product unbiased bases form a
\emph{nested} measurement (in the sense of \cite{TM4}), which means
that tracing out any of the parts of the POVMs the remaining measurements
are also product unbiased bases.

 Our main contribution consists of two examples that demonstrate the optimal allocation of quantum resources to be sound and tractable problems. Indeed, in our first example we show an explicit POVM $\widetilde{\mathbf{M}}=\left\{ \widetilde{M}_{x}\right\} _{x}$
for which the allocation $\mathcal{A}_{\mathcal{H}}\left[\widetilde{\mathbf{M}}\right]$
satisfies both optimality criteria $ \mathscr{C}_{1} $ and $\mathscr{C}_{2}$
for any $\mathcal{H}$ of qudits in the resource theory of incompatible
measurements. The elements of each $\widetilde{M}_{x}$ are given by the projectors
$\left|u_{\mathbf{a}}^{x=1}\right\rangle \left\langle u_{\mathbf{a}}^{x=1}\right|,\left|u_{\mathbf{b}}^{x=2}\right\rangle \left\langle u_{\mathbf{b}}^{x=2}\right|$,
onto the unbiased bases. Then we arrive at the following theorem:
\begin{theorem}
\label{thm:main} Consider an optimal allocation with
arbitrary hypergraph $\mathcal{H}$ of a given measurement $\mathbf{M}$ over a number $N$ of qudits of dimension
$d$ and with optimality criterion $\mathscr{C}$. If the criterion
$\mathscr{C}$ requires maximization of performance function $\Phi_{\mathscr{C}}$,
with allocation $\mathcal{A}_{\mathcal{H}}\left[\mathbf{M}\right]=\left\{ \mathcal{R}_{g}\left(\mathbf{M}_{\alpha}\right)\right\} _{\alpha\in\mathcal{E}_{\mathcal{H}}}$
defined by the generalized robustness $\mathcal{R}_{g}$, then the measurement
$\widetilde{\mathbf{M}}$ composed of two product unbiased bases is a feasible
optimal solution in the resource theory of incompatibility.
\end{theorem}

\textit{Proof:} First, we need to specify in which way $\mathcal{R}_{g}$ applies to every  possible measurement $\mathbf{M}_{\alpha}$ indexed by the hyperedge $\alpha$ of the hypergraph. We use equation (\ref{eq:robustness}) specified to the measurement $\mathbf{M}_{\alpha}$, where the mixture $(\mathbf{M}_{\alpha}+s\mathbf{N}_{\alpha})/(1+s)$ belongs to the set $\mathcal{F}_{\alpha} $ of compatible measurements only on the composed system $\alpha\in\mathcal{E}_{\mathcal{H}}$  and  $\mathbf{N}_{\alpha}$ is some measurement also on the systems in $\alpha$. 
This $\alpha$-wise application of $\mathcal{M}_{g}$ is meaningful since it should measure the advantage of $\mathbf{M}_{\alpha}$ over any compatible measurement of parties in $\alpha\in\mathcal{E}_{\mathcal{H}}$. Now, we remark the recent proof in \cite{RoB} that unbiased bases are optimal incompatible
measurements under the generalized robustness monotone $\mathcal{R}_{g}$
in any dimension $d$. Since, we can always find two product unbiased bases
to define $\widetilde{\mathbf{M}}$, then because product unbiased bases are
nested measurements the evaluation of $\mathcal{R}_{g}$ for each
$\alpha\in\mathcal{\mathcal{E}_{\mathcal{H}}}$ yields the maximal value,
hence for any monotone function $\Phi_{\mathscr{C}}$ of $\mathcal{A}_{\mathcal{H}}\left[\widetilde{\mathbf{M}}\right]=\left\{ \mathcal{R}_{g}\left(\widetilde{\mathbf{M}}_{\alpha}\right)\right\} _{\alpha\in\mathcal{E}_{\mathcal{H}}}$
\textendash{} such as $\Phi_{\mathscr{C}_{1}}$ or $\Phi_{\mathscr{C}_{2}}$\textendash{}
the $\widetilde{\mathbf{M}}$ measurement achieves the maximal value \textit{q.e.d}.

As an Illustration,  we present optimal allocations associated with the hypergraphs from Fig.\ref{fig:hypergraphs}. The maximal value $\Phi_{\mathscr{C}_{1}}^{max}$
achieved by $\Phi_{\mathscr{C}_{1}}\left(\mathcal{A}_{\mathcal{H}_{1}}\left[\widetilde{\mathbf{M}}\right]\right)$
is:
\begin{equation}
\Phi_{\mathscr{C}_{1}}^{max}=\log\left(\frac{d^{2}-1}{d^{2}+1}\right)+\log\left(\frac{d^{3/2}-1}{d^{3/2}+1}\right)
\end{equation}
Another example is the maximal value $\Phi_{\mathscr{C}_{2}}^{max}$
achieved by $\Phi_{\mathscr{C}_{2}}\left(\mathcal{A}_{\mathcal{H}_{2}}\left[\widetilde{\mathbf{M}}\right]\right)$
for each qudit pair of dimension $d$ is:
\begin{equation}
\Phi_{\mathscr{C}_{2}}^{max}=\pi_{\left\{ a,b\right\} }\left(\frac{d-1}{d+1}\right)+\left(\pi_{\left\{ a\right\} }+\pi_{\left\{ b\right\} }\right)\left(\frac{\sqrt{d}-1}{\sqrt{d}+1}\right)
\end{equation}
Since the resources in this particular resource theory are advantageous
for state discrimination, our result shows that the device which implements the optimal measurement $\widetilde{\mathbf{M}}$ also has optimal performance
for such a task in the presence of failures of any subsystem.
The scenario represented by $\mathcal{H}_{2}$ is the simplest case
in which \textquotedblleft at most a number $K$ of devices may fail\textquotedblright ,
which defines a problem of allocation relevant to the performance
of different engineering tasks \cite{ResOpt2}.

\subsection{Equitable allocation of nonlocality versus local contextuality}

The equitability criterion described in section 4 is relevant in contexts where there is some trade-off between resources distributed among agents. Such contexts are not strange in quantum information, for example, we can mention some cases such as extracting work from local states \cite{LocWork} versus obtaining thermodynamic work from shared correlations \cite{CorrWork0,CorrWork1} or obtaining random local states \cite{LocRand} versus sharing private bit states \cite{Qpriv}. For the tasks mentioned above, the cited articles quantify the appropriate resources with monotones based on entropic measures to which our theoretical framework applies. However, below we select cases of nonlocality versus local contextuality trade-offs due to their simplicity. Our choice of a specific case allows us to present a proof of concepts clearly and understandably.

 Nonlocality and contextuality, are well known resources for communication
and randomness amplification tasks \cite{Rand,Rand1,Comm,NonlocRes,Axiom}.
A monotone for these state resource theories is \cite{Axiom,NonlocRes}:
\begin{equation}
\mathcal{N}\left(\rho\right)=\sup_{\phi\in\Phi}\mathcal{I}\left(\phi\left[\rho\right]\right)-\mathcal{B}_{c}\label{eq:monotone}
\end{equation}
where $\mathcal{I}$ is the Bell correlation function, $\Phi$ is the set of free operations (see Appendix A) and $\mathcal{B}_{c}$ is the classical bound.
We will use the resource theories of nonlocality and contextuality
to exemplify a resource scenario in which the optimal equitability
criteria $\mathscr{C}_{3}$ is useful. In this scenario, a hypergraph $\mathcal{H}_{3}:\:\mathcal{V}_{\mathcal{H}_{3}}=\left\{ a,b\right\} ,\mathcal{E}_{\mathcal{H}_{3}}=\left\{ \left\{ a,b\right\} ,\left\{ b\right\} \right\} $ defines the allocation
of the resource state $\rho$.
The objective of party $b$ is to perform a task that improves with
the contextuality of his local state $\rho_{b}=\textrm{Tr}_{a}\left(\rho\right)$,
while parties $a,b$ should perform a task that requires nonlocality
of their bipartite state $\rho_{a,b}=\rho$. 

For our application of $\mathscr{C}_{3}$ is important the existence
of a fundamental trade-off between both resources \cite{Act} :
\begin{equation}
\mathcal{I}_{n}\left(\rho\right)+\mathcal{I}_{m}\left(\rho\right)\leq n+m-4\label{eq:tradeoff-1}
\end{equation}
Here $\mathcal{I}_{s}\left(\rho\right)$ stands for a cyclic correlation,
\begin{equation}
\mathcal{I}_{s}\left(\rho\right)=\sum_{k=1}^{s-1}\left\langle B_{k}B_{k+1}\right\rangle _{\rho}-\left\langle B_{s}B_{1}\right\rangle _{\rho},\label{eq:cyclineq}
\end{equation}
 where the average of observable $O=\left\{ O_{k}\right\} _{k}$ given
by $\left\langle O\right\rangle _{\rho}=\sum_{k}o_{k}Tr\left(O_{k}\rho\right)$, with $o_{k}\in \{-1,+1\}$ the numerical value of $O$ associated with $O_{k}$ and  $\mathcal{B}_{c}^{s}=s-2$ is the classical bound.
The correlation $\mathcal{I}_{s}\left(\rho\right)$ witnesses nonlocality
or contextuality depending on the kind of constraints satisfied by
the $\left\{ B_{1},...,B_{s}\right\} $ observables \cite{Act}.
Then, if we replace appropriately $\mathcal{I}_{s}$ and $\mathcal{B}_{c}^{s}$ in \eqref{eq:monotone} to define a monotone $\mathcal{N}_{s}\left(\rho\right)$  the inequality \eqref{eq:tradeoff-1} implies the resource relation:
\begin{eqnarray}
&\Lambda_{n}\left(\rho_{a,b}\right)&\mathcal{N}_{n}\left(\rho_{a,b}\right)+\Lambda_{m}\left(\rho_{b}\right)\mathcal{N}_{m}\left(\rho_{b}\right)\leq ...\nonumber\\
 &\Lambda_{n}\left(\rho_{a,b}\right)&\left[\mathcal{B}_{Q}^{n}-\mathcal{B}_{c}^{n}\right]+\Lambda_{m}\left(\rho_{b}\right)\left[\mathcal{B}_{Q}^{m}-\mathcal{B}_{c}^{m}\right]\,\,\, \label{eq:tradeoff2} \\
&\Lambda_{s}\left(\rho_{\alpha}\right)&=\frac{1+\textrm{sgn}\left(\mathcal{I}_{s}\left(\rho_{\alpha}\right)-\mathcal{B}_{c}^{s}\right)}{2}\nonumber 
\end{eqnarray}
    with $\mathcal{B}_{Q}^{s}$ is the quantum Tsirelson bound  (i.e. quantumly saturable) for $\mathcal{I}_{s}\left(\rho\right)$
and $\textrm{sgn}(\cdot)$ the sign function. From an allocation of resources
perspective the inequality \eqref{eq:tradeoff2} defines a knapsack
constraint like \eqref{eq:tradeoff}. The individual bounds (7) correspond here to $\mathcal{N}_n(\rho_{a,b})\leq \mathcal{B}^n_Q - \mathcal{B}^n_c$ and $\mathcal{N}_m(\rho_ b)\leq \mathcal{B}^m_Q- \mathcal{B}^m_c$
respectively. The optimal equitable solutions
in this case are simply:
\begin{eqnarray*}
\textrm{Solution 1:} &  & \left\{ \begin{array}{c}
\mathcal{N}_{n}\left(\rho_{a,b}\right)=\mathcal{B}_{Q}^{n}-\mathcal{B}_{c}^{n}\\
\\
\mathcal{N}_{m}\left(\rho_{b}\right)=0
\end{array}\right.\\
\\
\textrm{Solution 2:} &  & \left\{ \begin{array}{c}
\mathcal{N}_{n}\left(\rho_{a,b}\right)=0\\
\\
\mathcal{N}_{m}\left(\rho_{b}\right)=\mathcal{B}_{Q}^{m}-\mathcal{B}_{c}^{m}
\end{array}\right.
\end{eqnarray*}
If $\mathcal{B}_{Q}^{n}-\mathcal{B}_{c}^{n}>\mathcal{B}_{Q}^{m}-\mathcal{B}_{c}^{m}$
then the max-min algorithm will select Solution 1, since in this case $\mathcal{N}_{n}\left(\rho_{a,b}\right)>\mathcal{N}_{m}\left(\rho_{b}\right)$
is the choice that maximizes the overall amount of resources provided
by $\rho$. Conversely if $\mathcal{B}_{Q}^{n}-\mathcal{B}_{c}^{n}<\mathcal{B}_{Q}^{m}-\mathcal{B}_{c}^{m}$
we will obtain Solution 2 and finally, if $\mathcal{B}_{Q}^{n}-\mathcal{B}_{c}^{n}=\mathcal{B}_{Q}^{m}-\mathcal{B}_{c}^{m}$
both solution are acceptable, if only equitability is demanded.

A nontrivial scenario for optimal equitability arises in the case
of \emph{monogamy activation, }for example\emph{ }in the following
relationship \cite{Act}: 
\begin{equation}
\label{eq:tradeoffiq}
\mathcal{I}_{A}^{B}\left(\rho_{a,b}\right)+\mathcal{I}_{C}^{B}\left(\rho_{b,c}\right)+2\mathcal{I}_{5}\left(\rho_{b}\right)\leq14
\end{equation}
 
Here, $\mathcal{I}_{5}\left(\rho_{b}\right)$ is a contextual cycle correlation with $B_{k}$ observables as in \eqref{eq:cyclineq}, while
$\mathcal{I}_{A}^{B}\left(\rho_{a,b}\right),\mathcal{I}_{C}^{B}\left(\rho_{b,c}\right)$
stand for $I_{3322}$ inequalities \cite{Act}: 
\begin{multline}
    \mathcal{I}_{A}^{B}\left(\rho_{a,b}\right) = \left\langle B_{1}\right\rangle _{\rho}+\left\langle B_{4}\right\rangle _{\rho}+\left\langle A_{1}\right\rangle _{\rho}+\left\langle A_{2}\right\rangle _{\rho}-\left\langle B_{1}A_{1}\right\rangle _{\rho}\\
 -\left\langle B_{1}A_{2}\right\rangle _{\rho}-\left\langle B_{1}A_{3}\right\rangle _{\rho}-\left\langle B_{4}A_{1}\right\rangle _{\rho}-\left\langle B_{4}A_{2}\right\rangle _{\rho}\\
  +\left\langle B_{4}A_{3}\right\rangle _{\rho}-\left\langle B_{6}A_{1}\right\rangle _{\rho}+\left\langle B_{6}A_{2}\right\rangle _{\rho}\leq4
\end{multline} 

and analogously for $\mathcal{I}_{C}^{B}\left(\rho_{b,c}\right)$
by replacing each $A_{k}$ by $C_{k}$. Now, we study the allocation
scenario defined by $\mathcal{H}_{3}$, and with operational tasks
associated with $\mathcal{I}_{A}^{B}\left(\rho_{a,b}\right)$ and
$\mathcal{I}_{5}\left(\rho_{b}\right)$. In this case we assume $\rho_{b,c}$
to be resourceless, but with a fixed value $\mathcal{I}_{C}^{B}\left(\rho_{b,c}\right)=\lambda<4$.
Then, we can state an equitability problem defined by the constraints:
\begin{eqnarray}
\mathcal{N}_{A}^{B}\left(\rho_{a,b}\right)+2\mathcal{N}_{5}\left(\rho_{b}\right) & \leq & 4-\lambda\nonumber \\
0\leq\mathcal{N}_{A}^{B}\left(\rho_{a,b}\right) & \leq & \mathcal{B}_{Q}^{\gamma}-4\nonumber \\
0\leq\mathcal{N}_{5}\left(\rho_{b}\right) & \leq & \mathcal{B}_{Q}^{5}-3\label{eq:monactalloc}
\end{eqnarray}
The monotones $\mathcal{N}_{A}^{B}\left(\rho_{a,b}\right)$ and $\mathcal{N}_{5}\left(\rho_{b}\right)$
are defined analogously as $\mathcal{N}_{s}\left(\rho\right)$ in
\eqref{eq:monotone}, and $\mathcal{B}_{Q}^{\gamma}$, $\mathcal{B}_{Q}^{5}$
are the quantum bounds for $I_{3322}$ and $\mathcal{I}_{5}\left(\rho\right)$
respectively. In this case the solution is nontrivial and reads:
\begin{eqnarray}
\textrm{Solution 3:} &  & \left\{ \begin{array}{c}
\mathcal{N}_{A}^{B}\left(\rho_{a,b}\right)=\frac{\mu}{3}\\
\\
\mathcal{N}_{5}\left(\rho_{b}\right)=\frac{\mu}{3}
\end{array}\right.\label{eq:equitable}
\end{eqnarray}
with $\mu=4-\lambda$ (for the proof see Appendix A, section A.2).
As mentioned before, nonlocal and contextual resources are useful for
randomness amplification \cite{Rand,Rand1}. Therefore, an application
of the solutions to the problems presented is an equitable assignment
of security into quantum networks. In consequence, the above examples
show that optimal equitable allocations of resources are relevant for
concrete tasks in quantum information.

\section{Discussion}

In quantum information, a variety of approaches are used to determine
the practical value of quantum systems, such as cryptography, communication
capacity, computational complexity and thermodynamics. The common factor among these approaches is the search for quantum resources to benefit agents in performing specific tasks. Observed as a whole, these form an economy that requires tools to manage such quantum resources, optimal allocation being crucial. 

This article applies three different criteria of the optimal allocation of quantum resources and demonstrates solutions to the corresponding optimization problems. We propose the above results for applications in the interplay between quantum processing machines (computers) and a quantum communication network (see \cite{QNet, QNet1, QNet2}).

Indeed, our results contribute to the line of work proposed by S. Wehner et al. \cite{QuInt} for developing the Quantum Internet in six functionality driven stages. Precisely, Theorem 1 shows that there are multipartite quantum resources for communication tasks in prepare-and-measure networks with the optimally proportionally fair and reliable distribution. Such a result fits the type of advances desired for quantum networks in the second functionality stage of the Quantum Internet. 

Furthermore, our example of the optimal equitability criteria may be a starting point for analyzing network scenarios in the third stage proposed for Quantum Internet development (Entanglement distribution networks \cite{QuInt}). Indeed, nonlocality may be a self-testing benchmark for entanglement while contextuality may be related to coherence. The former may be related to communication between the nodes, and the latter - to quantum information processing at the nodes. If we have limited fault-tolerant resources for protecting both local and nonlocal quantumness, then the optimal equitability may help find a balance between local computation and delegation of tasks, especially if desired to be secure (see \cite{Kashefi-Pappa-2017}). Independently the very nonlocality at single computing node may be one day an  important resource itself (see \cite{Shallow-Networks-2018}). We leave these and related topics for further research.

In addition, the scheme we present opens the possibility of finding
applications for multi-party systems that nest resources symmetrically
\cite{Nest1}. In this sense, our results suggest that resource nesting may be potentially a tool for distributing resources according to specific criteria relevant to the corresponding tasks.

 In conclusion, it seems natural to expect that quantum resources will need soon the tools to allocate resources provided in this article.
It may be helpful at the level of designing quantum information processing protocols (or - possibly - even some practical experiments with limited quantum coherence effects). On the other hand, the methods and tools for allocating resources can contribute to the theoretical analysis of trade-offs between different quantum information resources. One can even expect that they can contribute to developing some form of the \emph{economy of quantum resources}. The administration of quantum networks -- such as the Quantum Internet -- could be one of this new economy's first challenge.

\section{Acknowledgements}
RS, TB and PH acknowledge support by the Foundation for Polish Science
(IRAP project, ICTQT, contract no.2018/MAB/5, co-financed by EU within Smart Growth Operational Programme). JC and KZ thank the support of National
Science Center in Poland under the grant numbers DEC-2019/35/O/ST2/01049  and  DEC-2015/18/A/ST2/00274 and by Foundation for Polish Science under the grant TEAM-NET project (contract no. POIR.04.04.00-00-17C1/18-00). We also thank A. de Oliveira Junior, Debashis Saha and Gabriel Muniz for advice and help with the design of figures.

\onecolumn
\appendix

\section{Appendix}
\subsection{Resource theories of nonlocality and contextuality}

The resource theories of nonlocality and contextuality usually consider
as objects boxes with at least two integer inputs $x_{1},x_{2}$ and
two integer outputs $a_{1},a_{2}$, which are characterized by the
conditional probabilities $P\left(a_{1},a_{2}\mid x_{1},x_{2}\right)$
called behaviours \cite{NonlocRes,Axiom}. In our research we consider
only quantum resources, hence every behaviour can be written as:
\begin{equation}
P\left(a_{1},a_{2}\mid x_{1},x_{2}\right)=Tr\left(\rho_{1,2}M_{a_{1}\mid x_{1}}^{\left(1\right)}\otimes M_{a_{2}\mid x_{2}}^{\left(2\right)}\right)
\end{equation}
For a bipartite state $\rho_{1,2}$ and measurements $M_{a_{1}\mid x_{1}}^{\left(1\right)}\otimes M_{a_{2}\mid x_{2}}^{\left(2\right)}$
for systems $1,2$. Moreover, in each case we consider fixed measurements,
such that in our study the resource theories of nonlocality and contextuality
can as well be considered resource theories of quantum states. Because
of this some examples of free operations for nonlocality and contextuality
can be local operations as well mixing of states. If in both cases
the set $\Phi$ is defined as the closure of free operations under
composition, the measure $\mathcal{N}_{s}\left(\rho\right)$ defined
in \eqref{eq:monotone} is a monotone: 
\begin{equation}
\mathcal{N}_{s}\left(\phi\left[\rho\right]\right)\leq\mathcal{N}_{s}\left(\rho\right)\:\forall\phi\in\Phi
\end{equation}
 due to the definition of supremum.

\subsection{Non-trivial solutions of allocation in the monogamy activation scenario}

Here we provide in more detail the possible equitable solutions to
the problem with knapsack constraints \eqref{eq:monactalloc}. First,
let's define the auxiliary variables $\mu=4-\lambda$, $\nu_{1}=\mathcal{B}_{Q}^{I_{3322}}-4$
and $\nu_{2}=\mathcal{B}_{Q}^{5}-3$. In Appendix C we show that $\nu_{1}\geq1$
and from reference \cite{Cyc} $v_{2}=0.9442$, hence $\nu_{1}>\nu_{2}.$
From the above and the partial order imposed by the equitability criteria
$\mathscr{C}_{3}$ we have: 
\begin{equation}
\mathcal{N}_{5}\left(\rho_{\left\{ b\right\} }\right)\leq\mathcal{N}_{A}^{B}\left(\rho_{\left\{ a,b\right\} }\right)\label{eq:ineqB}
\end{equation}
which means that in this scenario the advantage provided by nonlocality
of $\rho_{\left\{ a,b\right\} }$ is potentially greater than the
noncontextuality of $\rho_{\left\{ b\right\} }$, therefore has priority. Additionally we know (See ref \cite{Act, Cyc, Vertesi}) that all values for $\mathcal{I}_{5}\left(\rho_{\left\{ b\right\} }\right),\mathcal{I}_{A}^{B}\left(\rho_{\left\{ a,b\right\} }\right),\lambda$ satisfying the trade-off (\ref{eq:tradeoffiq}) and Tsirelson bounds to be achievable for some $\rho_{ a,b,c}$, then the supreme in each step of the optimal equitability criterion can be replaced by a maximization. Later, because the optimization is over a convex set -- of $\mathcal{N}_{5}\left(\rho_{\left\{ b\right\} }\right),\mathcal{N}_{A}^{B}\left(\rho_{ a,b}\right)$-- the solution must lie
at a boundary. A solution in the boundary $\mathcal{N}_{5}\left(\rho_{\left\{ b\right\} }\right)=\nu_{2}$
will violate the requirements \eqref{eq:monactalloc} or \eqref{eq:ineqB}
because $\mu\leq2\nu_{2}$, on the other hand a solution with $\mathcal{N}_{A}^{B}\left(\rho_{\left\{ a,b\right\} }\right)=\nu_{1}$,
implies $\mathcal{N}_{5}\left(\rho_{\left\{ b\right\} }\right)=0$
which is actually a minimum. Then, to find a maximum for $\mathcal{N}_{5}\left(\rho_{\left\{ b\right\} }\right)$
under the constraints, we search at the boundary:
\begin{eqnarray}
\mathcal{N}_{A}^{B}\left(\rho_{\left\{ a,b\right\} }\right)+2\mathcal{N}_{5}\left(\rho_{\left\{ b\right\} }\right) & = & \mu\nonumber \\
\Rightarrow\mathcal{N}_{A}^{B}\left(\rho_{\left\{ a,b\right\} }\right) & = & \mu-2\mathcal{N}_{5}\left(\rho_{\left\{ b\right\} }\right)\label{eq:boundary}
\end{eqnarray}
Replacing \eqref{eq:boundary} in \eqref{eq:ineqB} we obtain: 
\begin{equation}
\mathcal{N}_{5}\left(\rho_{\left\{ b\right\} }\right)\leq\frac{\mu}{3}<\nu_{2}
\end{equation}
Since the lower bound of $\mathcal{N}_{A}^{B}\left(\rho_{\left\{ a,b\right\} }\right)$
is zero, the maximum value that $\mathcal{N}_{5}\left(\rho_{\left\{ b\right\} }\right)$
can achieve is $\mu/3$ in which case we have the equitable Solution
3 of main text. The solution \eqref{eq:equitable} maximizes the amount
of resources under constraints \eqref{eq:monactalloc} and minimizes
the difference between $\mathcal{N}_{A}^{B}\left(\rho_{\left\{ a,b\right\} }\right)$
and $\mathcal{N}_{5}\left(\rho_{\left\{ b\right\} }\right)$.

\subsection{The quantum bound of $I_{3322}$}





The Bell operator $\mathbbm{B}_{Q}^{v}$ of the $I_{3322}$ inequality was introduced in reference \cite{Vertesi}:

\begin{eqnarray}
 &  & \mathbbm{B}_{Q}^{v}=-A_{2}\otimes \mathbbm{I}-\mathbbm{I}\otimes B_{1}-2\mathbbm{I}\otimes B_{4}+A_{1}\otimes B_{1}\nonumber \\
 &  & +A_{1}\otimes B_{4}+A_{2}\otimes B_{1}+A_{2}\otimes B_{4}-A_{1}\otimes B_{6}\nonumber \\
 &  & +A_{2}\otimes B_{6}-A_{3}\otimes B_{1}+A_{3}\otimes B_{4}\label{eq:vertesi}
\end{eqnarray}
in terms of binary operators $\left\{ A_{1},A_{2},A_{3},B_{1},B_{4},B_{6}\right\} $
with outcomes $0$ and $1$, 
\begin{eqnarray*}
A_{i} & = & (1)A_{i}^{1}+(0)A_{i}^{\perp}\:i\in\left\{ 1,2,3\right\} \\
B_{j} & = & (1)B_{j}^{1}+(0)B_{j}^{\perp}\:j\in\left\{ 1,4,6\right\} 
\end{eqnarray*}
However, in this article we use binary operators with outcomes $-1$
and $1$,
\begin{eqnarray*}
A_{i}^{\prime} & = & (1)A_{i}^{1}+(-1)A_{i}^{\perp}\:i\in\left\{ 1,2,3\right\} \\
B_{j}^{\prime} & = & (1)B_{j}^{1}+(-1)B_{j}^{\perp}\:j\in\left\{ 1,4,6\right\} 
\end{eqnarray*}
Then, in this appendix we will show how to transform $\mathbbm{B}_{Q}^{v}$
into an operator $\mathbbm{B}_{Q}^{\gamma}$ in terms of binary operators
with outcomes $-1$ and $1$. First, note that:
\begin{eqnarray}
A_{i}^{1}+A_{i}^{\perp} & = & \mathbbm{I}\label{eq:id1-1}\\
B_{j}^{1}+B_{j}^{\perp} & = & \mathbbm{I}\label{eq:id2-1}
\end{eqnarray}
for all $i\in\left\{ 1,2,3\right\} $ and $j\in\left\{ 1,4,6\right\} $
respectively. In consequence:
\begin{eqnarray}
A_{i}=A_{i}^{1} & = & \frac{1}{2}\left(\mathbbm{I}+A_{i}^{\prime}\right)\label{eq:id3}\\
B_{j}=B_{j}^{1} & = & \frac{1}{2}\left(\mathbbm{I}+B_{j}^{\prime}\right)\label{eq:id4}
\end{eqnarray}
Second, we apply \eqref{eq:id3}, \eqref{eq:id4} to replace each $A_{i}$ and $B_{j}$ in the definition of $\mathbbm{B}_{Q}^{v}$ we obtain: 


%
\begin{eqnarray*}
 &  & \mathbbm{B}_{Q}^{v} =-\left(\frac{\mathbbm{I}\otimes \mathbbm{I}+A'_{2}\otimes \mathbbm{I}}{2}\right)-\left(\frac{\mathbbm{I}\otimes \mathbbm{I}+\mathbbm{I}\otimes B'_{1}}{2}\right)\\
 &  & -2\left(\frac{\mathbbm{I}\otimes \mathbbm{I}+\mathbbm{I}\otimes B'_{4}}{2}\right)+\left(\frac{\mathbbm{I}\otimes \mathbbm{I}+A'_{1}\otimes \mathbbm{I}+\mathbbm{I}\otimes B'_{1}+A'_{1}\otimes B'_{1}}{4}\right)\\
 &  & +\left(\frac{\mathbbm{I}\otimes \mathbbm{I}+A'_{1}\otimes \mathbbm{I}+\mathbbm{I}\otimes B'_{4}+A'_{1}\otimes B'_{4} }{4}\right)+\left(\frac{\mathbbm{I}\otimes \mathbbm{I}+A'_{2}\otimes \mathbbm{I}+ \mathbbm{I}\otimes B'_{1}+A'_{2}\otimes B'_{1}}{4}\right)\\
 &  & +\left(\frac{\mathbbm{I}\otimes \mathbbm{I}+A'_{2}\otimes \mathbbm{I}+\mathbbm{I}\otimes B'_{4}+A'_{2}\otimes B'_{4} }{4}\right)-\left(\frac{\mathbbm{I}\otimes \mathbbm{I}+A'_{1}\otimes \mathbbm{I}+\mathbbm{I}\otimes B'_{6}+A'_{1}\otimes B'_{6} }{4}\right)\\
 &  & +\left(\frac{\mathbbm{I}\otimes \mathbbm{I}+A'_{2}\otimes \mathbbm{I}+\mathbbm{I}\otimes B'_{6}+A'_{2}\otimes B'_{6} }{4}\right)-\left(\frac{\mathbbm{I}\otimes \mathbbm{I}+A'_{3}\otimes \mathbbm{I}+\mathbbm{I}\otimes B'_{1}+A'_{3}\otimes B'_{1}}{4}\right)\\
 &  & +\left(\frac{\mathbbm{I}\otimes \mathbbm{I}+A'_{3}\otimes \mathbbm{I}+\mathbbm{I}\otimes B'_{4}+A'_{3}\otimes B'_{4} }{4}\right)
\end{eqnarray*}

After some algebraic simplifications we obtain:
\begin{eqnarray}
  4\mathbbm{B}_{Q}^{v}+4 \mathbbm{I}\otimes \mathbbm{I} =A'_{1}\otimes \mathbbm{I}+A'_{2}\otimes \mathbbm{I}-\mathbbm{I}\otimes B'_{1}-\mathbbm{I} \otimes B'_{4}+A'_{1}\otimes B'_{1} +A'_{1}\otimes B'_{4}\nonumber\\
  +A'_{2}\otimes B'_{1} +A'_{2}\otimes B'_{4}-A'_{1}\otimes B'_{6} +A'_{2}\otimes B'_{6}-A'_{3}\otimes B'_{1} +A'_{3}\otimes B'_{4}\label{eq:id5}
\end{eqnarray}
If now we use observables $B_{1}^{\prime\prime}=-B'_{1}$, $B_{4}^{\prime\prime}=-B'_{4}$
and $A_{3}^{\prime\prime}=-A'_{3}$ which are just re-labeling of outcomes for the observables $B'_{1},B'_{4},A'_{3}$ we have the following relation:
\begin{eqnarray}\label{eq:bV}
 4 \mathbbm{B}_{Q}^{v}+4\mathbbm{I} \otimes \mathbbm{I}= A'_{1}\otimes \mathbbm{I}+A'_{2}\otimes \mathbbm{I}+\mathbbm{I}\otimes B''_{1} +\mathbbm{I}\otimes B''_{4} -A'_{1}\otimes B_{1}^{''} - A'_{1}\otimes B_{4}''\nonumber \\
 - A'_{2}\otimes B_{1}^{''} -A'_{2}\otimes B_{4}^{''} - A'_{1}\otimes B_{6}' +A'_{2}\otimes B'_{6} \nonumber  -A''_{3}\otimes B_{1}^{''} + A_{3}^{''}\otimes B''_{4} 
\end{eqnarray}
Now, in the article we used the alternative form of $I_{3322}$ in terms of binary operators with outputs ${+1,-1}$ whose corresponding Bell operator is:
\begin{eqnarray}
& &\mathbbm{B}_{Q}^{\gamma}=  A'_{1}\otimes \mathbbm{I}+A'_{2}\otimes \mathbbm{I}+\mathbbm{I}\otimes B''_{1} +\mathbbm{I}\otimes B''_{4} -A'_{1}\otimes B_{1}^{''} - A'_{1}\otimes B_{4}''\nonumber \\
 &  & - A'_{2}\otimes B_{1}^{''} -A'_{2}\otimes B_{4}^{''} - A'_{1}\otimes B_{6}' +A'_{2}\otimes B'_{6} \nonumber  -A''_{3}\otimes B_{1}^{''} + A_{3}^{''}\otimes B''_{4}  \label{eq:bS}
\end{eqnarray}
By proper identification of terms in \eqref{eq:bV} and \eqref{eq:bS},
we obtain the operator identity: 
\begin{equation}
\mathbbm{B}_{Q}^{\gamma}=4\mathbbm{B}_{Q}^{v}+4 \mathbbm{I}\otimes \mathbbm{I} \label{eq:OpId}
\end{equation}
Additionally, from the definition of Tsirelson bounds $\mathcal{B}_{Q} ^{v}$ and $\mathcal{B}_{Q} ^{\gamma}  $ we have:
\begin{equation}
  \sup_{\substack{{\rho \in \mathcal{S}_2} \\ {A_1,A_2,A_3,B_1,B_2,B_3 \in \mathcal{L}(\mathbbm{C}^2)} }} \mathrm{Tr}(\mathbbm{B}_{Q}^{v}\rho) = \mathcal{B}_{Q} ^{v} 
\end{equation}
\begin{equation}
  \sup_{\substack{{\rho \in \mathcal{S}_2} \\ {A_1,A_2,A_3,B_1,B_2,B_3 \in \mathcal{L}(\mathbbm{C}^2)} }} \mathrm{Tr}(\mathbbm{B}_{Q}^{\gamma}\rho) = \mathcal{B}_{Q} ^{\gamma} 
\end{equation}
but in reference \cite{Vertesi} is shown that a lower bound for $\mathcal{B}_{Q}^{v}$
is $0.25$ and an upper bound is $0.25085...$, then  identity \eqref{eq:OpId} and Tsirelson bound definition implies:
\begin{equation}
1\leq\mathcal{B}_{Q}^{\gamma}-4\leq1.0034
\end{equation}

which are the bounds used in section A.2 of the Appendix.

\bibliographystyle{plain}

\end{document}